\documentclass[11pt,twoside]{article}
\usepackage{asp2004}
\usepackage{psfig}
\usepackage{epsf}
\usepackage{graphics}
\usepackage{lscape}
\def\edcomment#1{\iffalse\marginpar{\raggedright\sl#1\/}\else\relax\fi}
\reversemarginpar

\begin{document}
\title{SNLS: Overview and High-z Spectroscopy}
\author{D. Andrew Howell for the SNLS Collaboration}
\affil{Department of Astronomy and Astrophysics, The University of
  Toronto, 60 St. George Street, Toronto, ON, Canada}

\begin{abstract}
  The Supernova Legacy Survey (SNLS)\footnote{For a complete author
    list and candidates see http://cfht.hawaii.edu/SNLS} will discover
  and obtain $g' r' i' z'$ lightcurves for more than 700
  spectroscopically confirmed SNe Ia $(0.3<z<0.9)$ to differentiate
  between competing models for Dark Energy.  We fit the multicolor
  lightcurves of the candidates to determine which are likely SNe Ia
  and send them for follow-up spectroscopy to the Keck, VLT, and
  Gemini telescopes.  Here we show the results from Gemini, where we
  send most of our highest redshift $(0.6<z<0.9)$ targets.

\end{abstract}
\thispagestyle{plain}

\section{Lightcurves}
As a part of the Canada-France-Hawaii Telescope Legacy Survey
(CFHTLS), the SNLS uses a rolling search strategy to discover and
monitor SNe in four one degree fields over the course of five years.  
SNe are observed in $g' r' i' z'$, allowing unprecedented
measurement of reddening and SN intrinsic colors.  Observations are made
every 2-3 rest frame days during dark and gray time.  A typical run
includes 10 observing blocks of 5 visits to each of two deep fields.
Because we always observe the same fields, every
visit brings new SN discoveries and simultaneously adds points to existing
lightcurves.  See the contribution of Sullivan et al. to these
proceedings for an overview of the SNLS.

\section{Spectroscopy}
For the most promising candidates, spectroscopy on 8-10m telescopes
must be obtained to determined the redshift and SN type.
We prescreen candidates -- early data are fit with a Ia 
lightcurve and poor fits (core-collapse SNe or 
AGN) are rejected.  For good fits (likely SNe Ia), the redshift and 
phase are estimated, allowing optimal scheduling of observations and
instrument setup.

Spectroscopy of candidate SNe is carried out on Gemini, Keck and VLT.
The highest redshift targets are observed at Gemini N and S with GMOS
(Hook et al. 2004), where we make use of the Nod and Shuffle mode to 
virtually eliminate the systematic errors associated with the
subtraction of sky lines in the red.  Gemini also observes some lower
redshift targets in classical mode in the D3 (Extended Groth Strip)
field, which cannot be seen by VLT.  All GMOS observations use the
R400 grating, 0.75\arcsec\ slit, and either the 680 or 720nm central
wavelength setup.  The CCD is binned $2 \times 2$.

\section{Identification}
Spectra are fit with a $\chi^2$ matching program (Howell \& Wang 2002)
against a library of SNe templates.  Template host galaxy light is 
also fit and subtracted, and the redshift of the SN is estimated.
This permits a determination of the SN type even when it is
buried in a host galaxy.  An expert must still make the final
decision, but the quantitative comparison of 
hundreds of SN spectra offers robustness, accuracy, and speed 
advantages over unaided human classification.  

\section{Results}
At Gemini, in semesters 2003B and 2004A we observed 48 targets, 
29 of which are consistent with a SN Ia classification.
We show some example spectra in Figure 1.

\acknowledgements
The SNLS collaboration acknowledges the
CFHT Queue Service Observing team, the agencies CNRS/IN2P3,
CNRS/INSU and CEA in France, and NSERC and CIAR in Canada.  
Based on observations obtained at the Gemini Observatory, and with
Mega Prime, a joint project of CFHT, CEA/DAPNIA, and HIA.

\begin{center}

\begin{minipage}[h]{6.5in}

\scalebox{0.25}{
\includegraphics{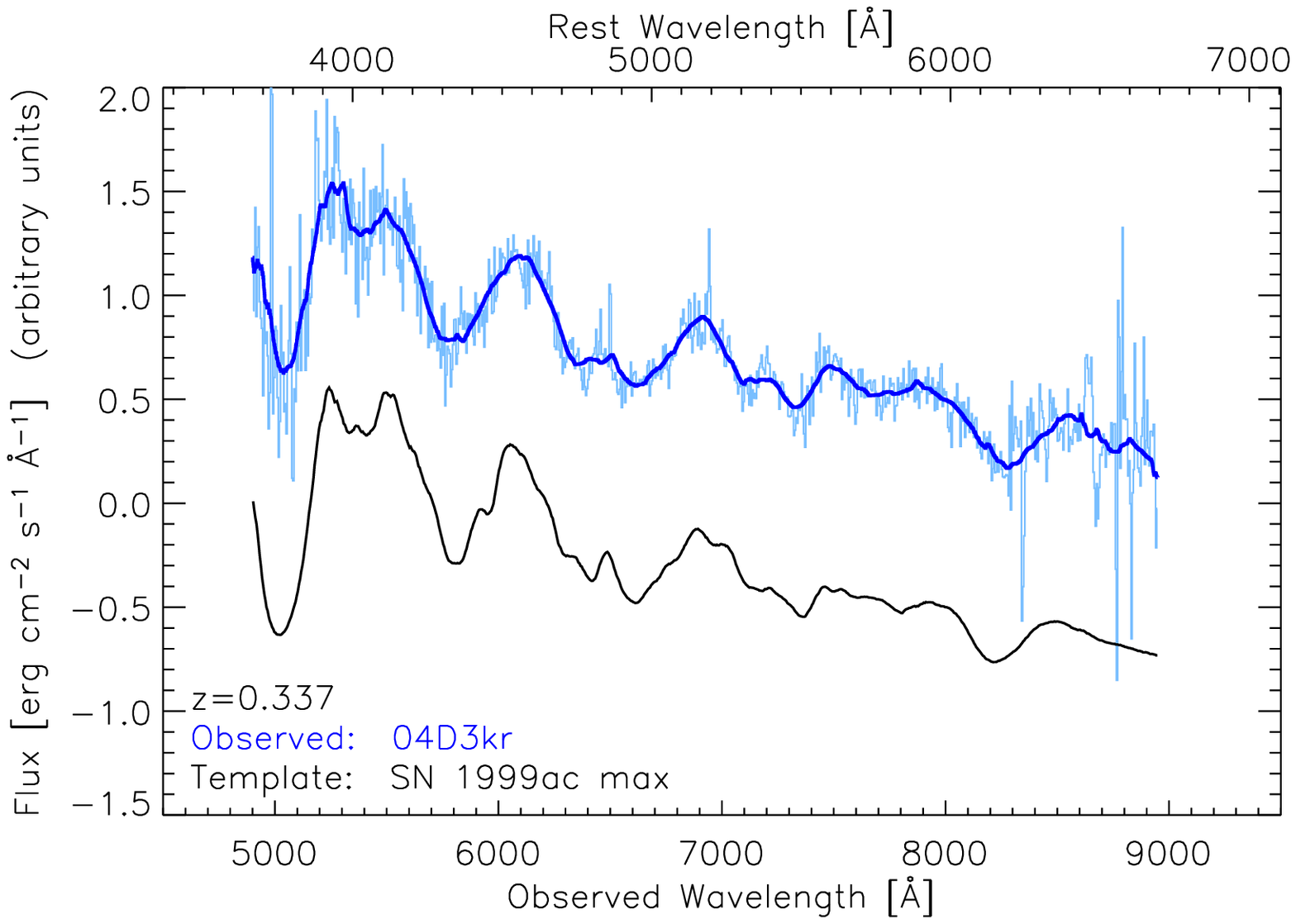}}
\scalebox{0.25}{
\includegraphics{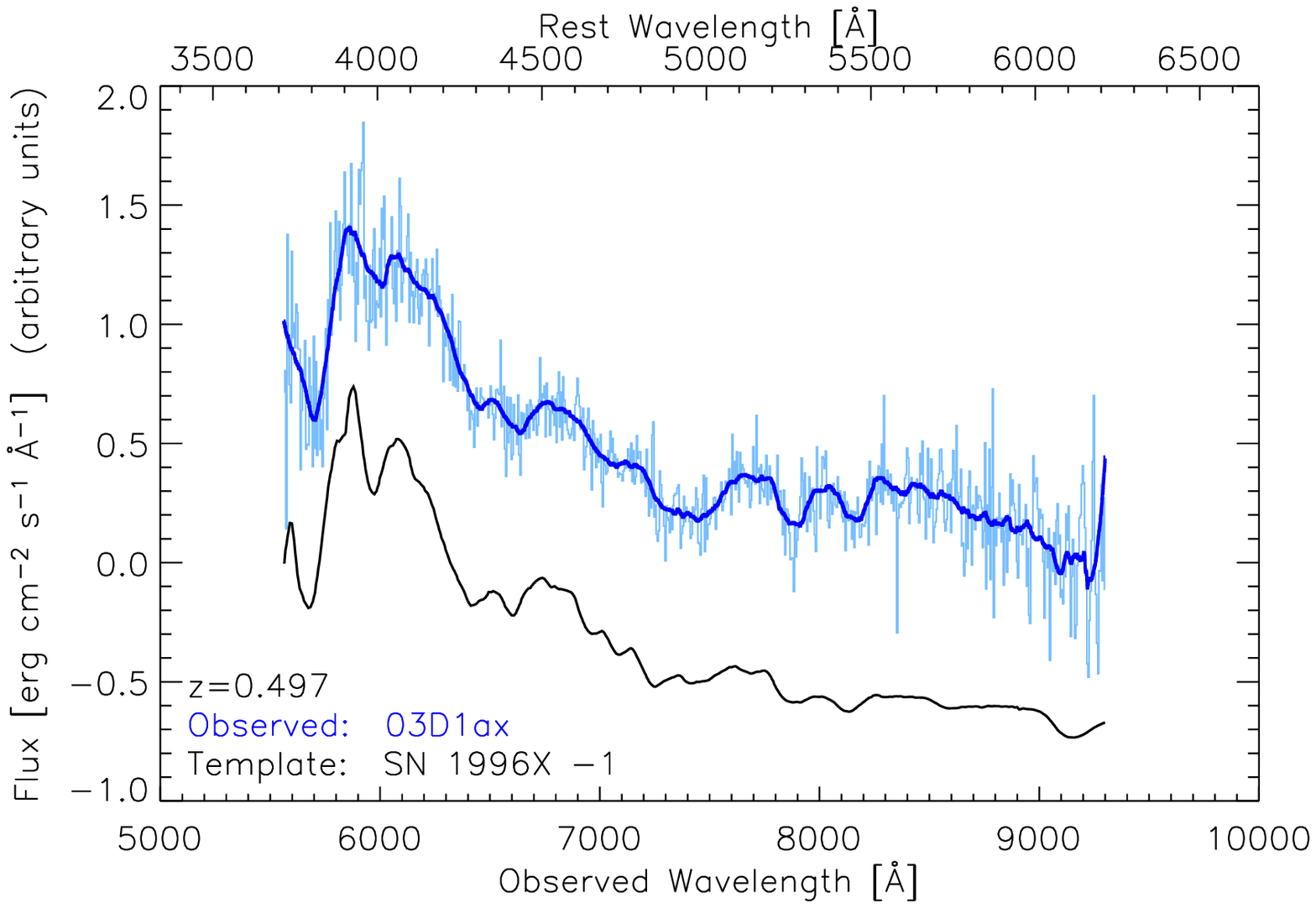}}
\scalebox{0.25}{
\includegraphics{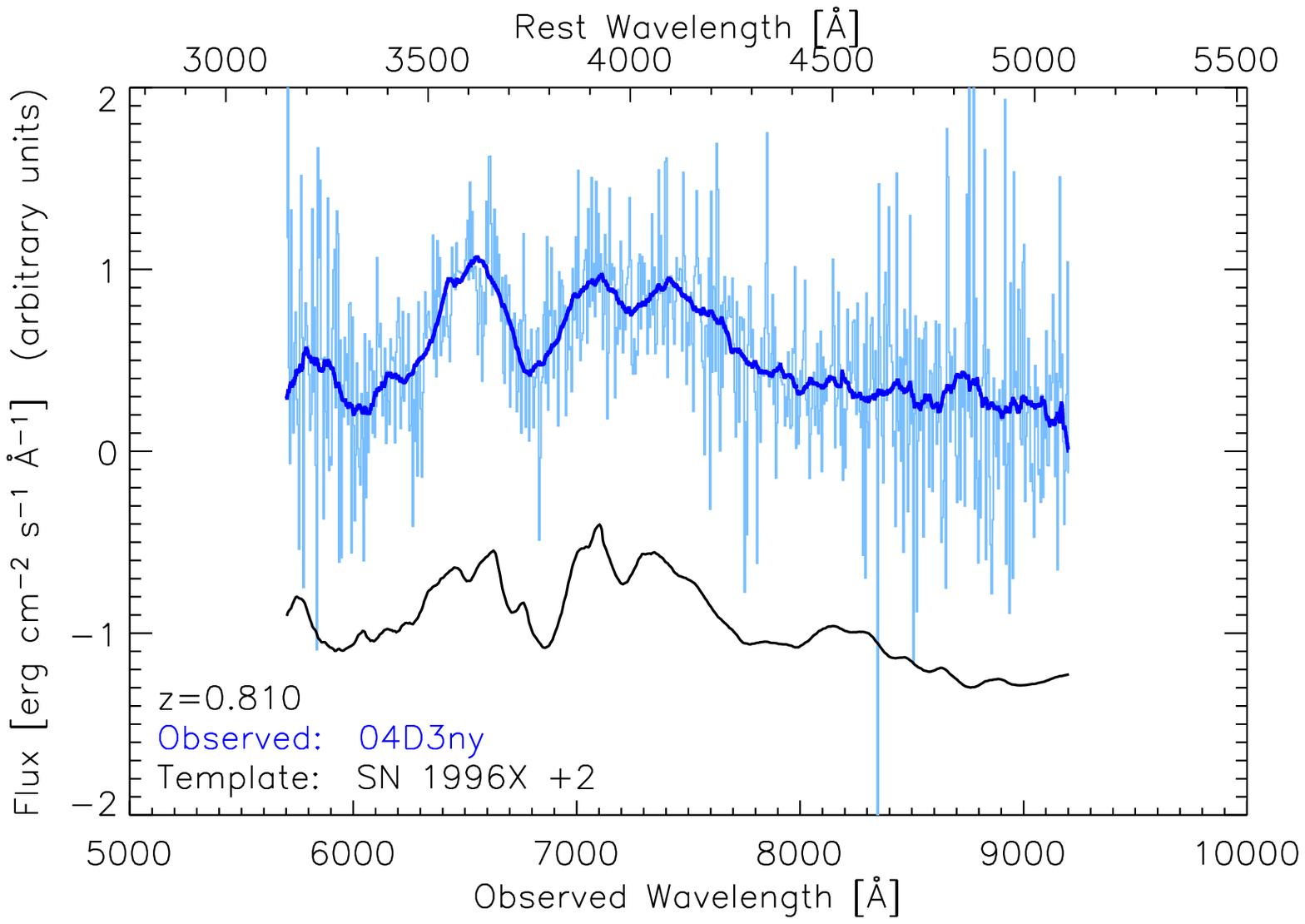}}

\vspace{-0.07in}

\scalebox{0.25}{
\includegraphics{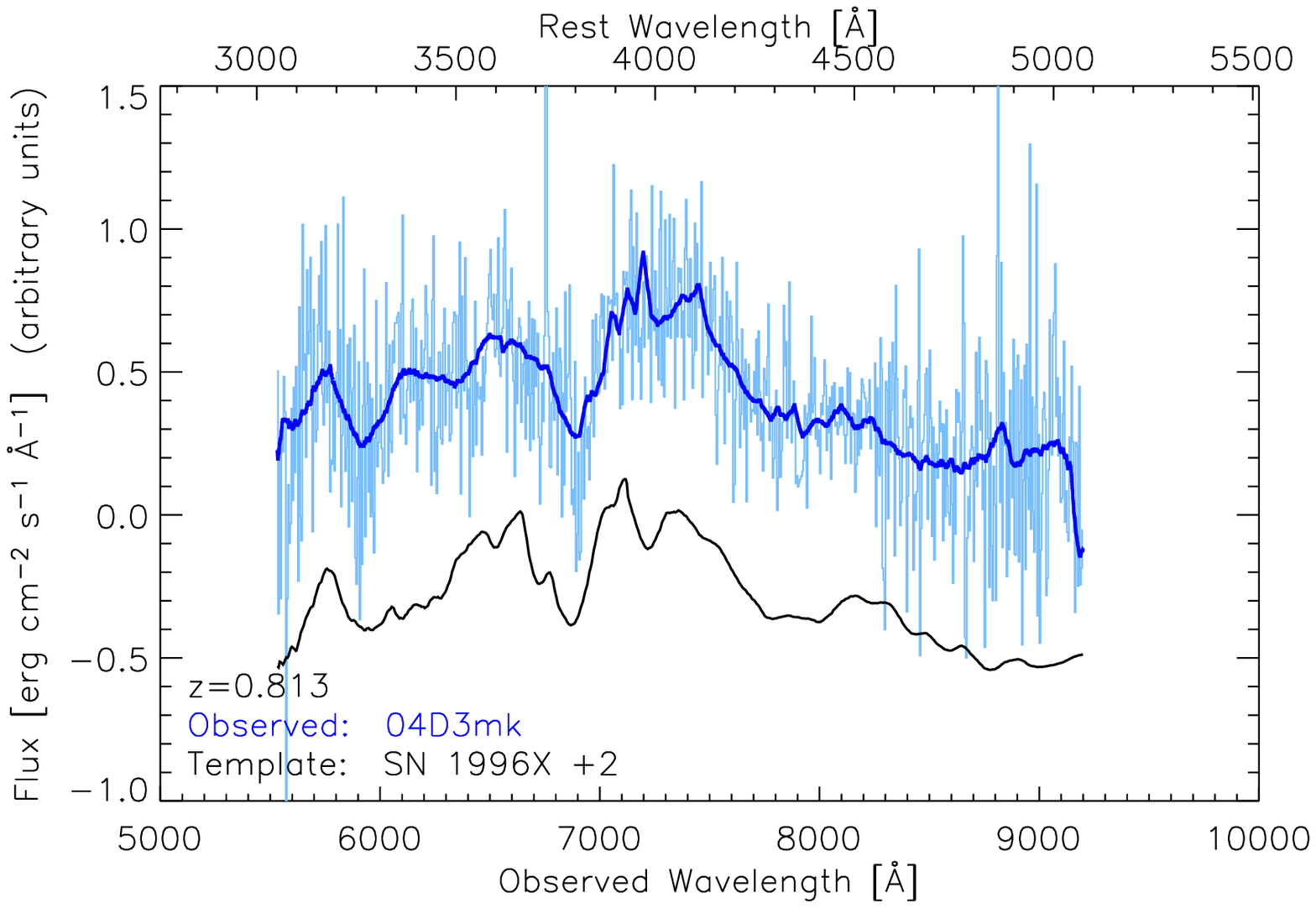}}
\scalebox{0.25}{
\includegraphics{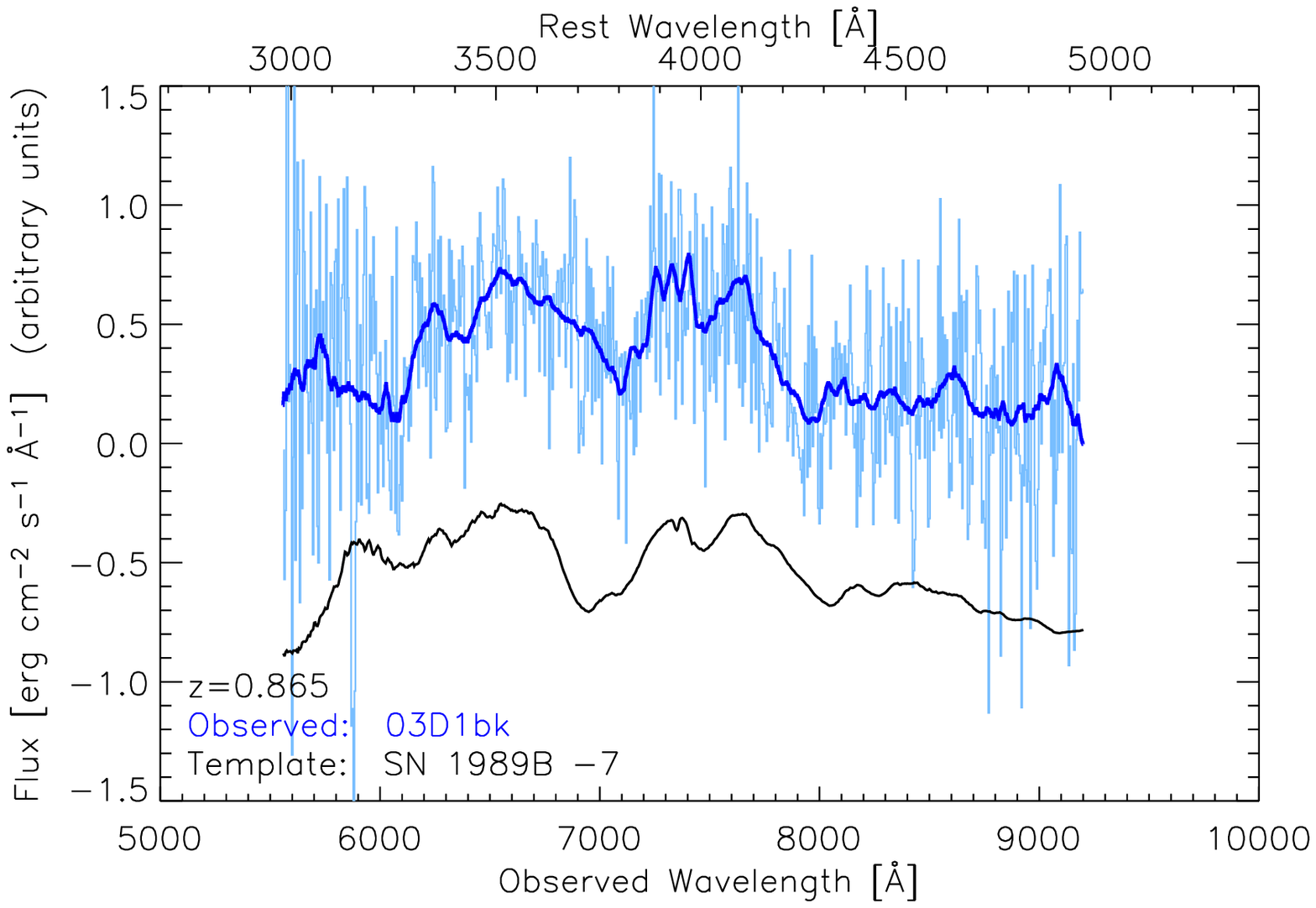}}
\scalebox{0.25}{
\includegraphics{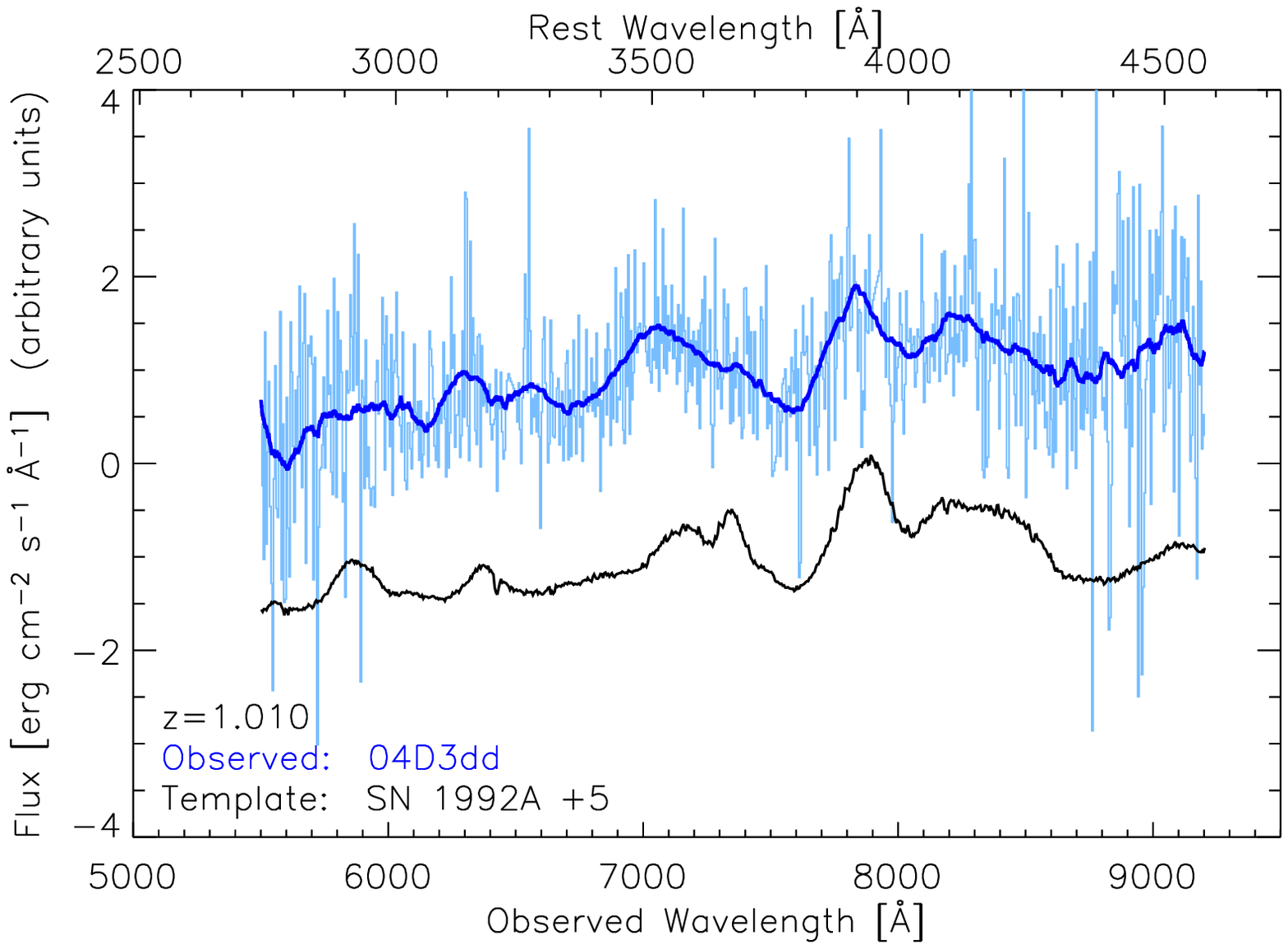}}

\end{minipage}

\end{center}

{\small \textbf{\textsf{Figure 1: }} Example Gemini spectra of
  confirmed Type Ia SNLS SNe from 2003B/04A.  We show data after host galaxy 
  subtraction (if necessary), re-binned to 5\AA , with smoothed data 
  overplotted.  The best-fitting SN template is shown
  underneath.}

\end{document}